\documentclass[10pt,a4paper,twoside]{article}
\usepackage{epsfig}
\usepackage{baltlat6}
\usepackage{array}
\usepackage{here}
\usepackage{natbib}
\begin{document}
\ \
\vspace{0.5mm}
\setcounter{page}{1}

\titlehead{Baltic Astronomy, vol.\,23, 2014}

\titleb{COSMIC-RAY INDUCED DIFFUSION IN INTERSTELLAR ICES}

\begin{authorl}
\authorb{Juris Kalv\=ans}{}
\end{authorl}

\begin{addressl}
\addressb{}{Engineering Research Institute "Ventspils International Radio Astronomy Center" of Ventspils University College, Inzenieru 101, Ventspils, Latvia, LV-3601}
\end{addressl}

\submitb{Received: 2014 April 16}

\begin{summary}

Cosmic rays are able to heat whole interstellar dust grains. This may enhance molecule mobility in icy mantles that have accumulated on the grains in dark cloud cores.

A three-phase astrochemical model was used to investigate molecule mobility in interstellar ices. Specifically, diffusion through pores in ice between the subsurface mantle and outer surface, assisted by whole-grain heating, was considered.

It was found that the pores can serve as an efficient transport route for light species. The diffusion of chemical radicals from the mantle to the outer surface has the most notable effect. These species accumulate in the mantle because of photodissociation by cosmic-ray induced photons. The faster diffusion of hydrogen within the warm ice enhances the hydrogenation of radicals on pore surfaces. The overall result of whole grain heating-induced radial diffusion in ice is higher abundances for ice species whose synthesis involve light radicals. Examples of stable species synthesized this way include complex organic molecules, OCS, H$_2$O$_2$, and cyanoplyynes.

\end{summary}

\begin{keywords} astrochemistry -- molecular processes -- ISM: clouds -- ISM: molecules -- ISM: cosmic rays
\end{keywords}

\resthead{Cosmic-ray induced diffusion in interstellar ices}
{J. Kalv\=ans}

\sectionb{1}{INTRODUCTION}

Dark cores in giant interstellar molecular clouds are the densest and coldest regions of interstellar medium. Many of these cores represent the birth sites of stars. Most of the chemical elements heavier than helium (hereafter referred to as 'metals') are in the form of dust grains or "dirty ices" -- molecular species frozen onto the interstellar grains.

Interstellar ices are affected by ionizing irradiation that is capable to induce molecule dissociation \citep{Ruffle01a}. Interstellar UV radiation drives ice photoprocessing in diffuse clouds, while cosmic-ray-induced photons \citep{Prasad83} is the likely molecule destruction agent in dark cloud cores.

A separate process that affects interstellar ices is whole-grain heating by heavy cosmic-ray (CR) nuclei \citep{Leger85}. The evaporation from such warm grains affects gas-grain chemistry \citep{Hasegawa93a}, while the enhanced rate of surface diffusion may promote the formation of complex species in ice \citep{Reboussin14}.

The aim of the present study is to investigate the possible chemical significance of radial diffusion of molecules through pores in ice, induced by whole-grain heating. This is not to be confused with \textit{lateral} surface diffusion, studied by \citet{Reboussin14}. With radial diffusion we mean the exchange of species between the outer surface and the mantle below it. The details of this principle are explained in Sect.~2.

In order to study the radial diffusion process, a three-phase (gas, surface, and bulk ice) model is required. While modeling of surface reactions dates back to 1970s \citep[e.g.][]{Watson72}, the investigation of active subsurface ice chemistry is relatively recent. \citet{Cuppen07} and \citet{Kalvans10} have published the first such models with the Monte Carlo technique and the rate equations method, respectively. Detailed ice mantle modeling provides important information about several astrochemistry problems, e.g., ice growth and structure, hydrogenation and oxidation of CO on grains, and production of complex organic molecules (COM) \citep{Cuppen09,Kalvans13a,Garrod13a,Garrod13b}.

The presented model considers subsurface ice mantle as a fully mixed single phase that contains cavities in it. These cavities -- representing a low-level subsurface ice porosity -- are the special feature of the model \citet{Kalvans10}. The aim of this research is to perform a study how the chemical composition of ice will change, if the pores are opened and connect to the surface. This is assumed to occur when ice structure becomes unsteady during the whole-grain heating events.

\sectionb{2}{THE MODEL}
\subsectionb{2.1}{Model system}

The model employed in the present study has been developed by \citet{Kalvans10,Kalvans13a}. It was transferred to the Heidelberg astrochemical code ALCHEMIC by \citet{Kalvans13c}. ALCHEMIC \citep{Semenov10} was kindly provided by Dmitry Semenov. The new model is dubbed `Alchemic-Venta'.

The ``traditional'' three phases, which are usually considered in astrochemical calculations (gas, grain surface, and ice mantle), are supplemented by an additional phase -- the surface of closed cavities (pores, cracks, tunnels) within the ice mantle.

The basic assumptions in the model follow. Gas-phase species (Sect.~2.3) accrete onto dust grains, forming an initially porous layer (the outer surface, \textit{S}). These surface species can be desorbed by evaporation, photodesorption (Sect.~2.4), and CR-induced processes. The porous outer layer is then compacted by cosmic rays and other effects, which form a compact mantle (\textit{M}) that still contains some sealed cavities (\textit{C}, Sect.~2.5). Reactions on outer and inner ice surfaces proceed, and ice molecules can be dissociated by interstellar ultraviolet photons or cosmic-ray induced photons (Sect.~2.6). Light hydrogen species H and H$_2$ diffuse (or hop) between the surface, mantle, and cavity phases (Sect.~2.5). Additionally, as a means for ice mixing, a fourth phase is considered -- a fluid that exists for a short time in a locally heated cylinder along the path of a heavy CR particle passing through the grain (Sect.~2.5). This is referred to as the `hot-soup' phase (\textit{H}). 

The molecules can change phases by the spot heating process \citep{Leger85}. When a heavy CR particle (iron nuclei, Fe-CR) impacts, a hot cylinder is formed for a very short time. It was assumed that the ice in the cylinder becomes fully mixed and then freezes again. Transfer of species between the surface, cavity, and mantle phases occurs in the cylinder, in accordance with their respective proportions, which are assumed constant.

Whole grain heating is induced by Fe-CR (after the spot-heating event) and lasts for $10^{-5}$s. During this time, evaporation of surface species occur, as well as enhanced diffusion of ice species. It was assumed that the ice becomes unsteady and the ``walls'' that separate cavities and the outer surface become leaky, and an exchange of surface molecules occurs between the two phases (Sect.~2.6). 

In the following sections it is explained in detail how these processes are reflected in the model, or referred to other sources. For processes, where transition between the gas, outer-surface, cavity-surface, mantle, or hot-soup phases occur, the initial phase is abbreviated \textit{f}0 and the final phase is \textit{f}. The phases and their transitions are explained below.

\subsectionb{2.2}{Physical conditions}
\label{cond}

The model is used to explore chemical evolution of a hypothetical prestellar core with a constant density ($n_\mathrm{H}=2\times10^4$cm$^{-3}$) and temperature (10K). The total integration time \textit{t} was taken 3Myr, however 1Myr was always used as a `standard time' when interpreting the results.

Elemental abundances are taken the same as in \citet{Kalvans13a}. Deuterium chemistry was not included. The abundances for three additional elements Si, P, and Cl were adopted from \citet{Jenkins09} with $F_* = 1$. Initially, all the metal elements were assumed to be in gas-phase atomic form. H$_2$ contains most of hydrogen  molecules, with atomic H relative abundance of only 0.002. The grains are initially neutral.

A standard cosmic-ray ionization rate of $1.3\times10^{-17}$s$^{-1}$ \citep{Herbst73} was adopted. The extinction of the interstellar UV radiation field $A_V$ was taken to be 20mag, more consistent for a large molecular cloud than an isolated dense core. This is the same value employed by \citet{Kalvans13a}. The grains were assumed to be of uniform size, with a radius $a=10^{-5}$cm. About 100 monolayers of ice species accumulate onto the grains during cloud evolution. It was assumed that each monolayer has a thickness of $b_\mathrm{ML}=3.7 \times 10^{-8}$cm. Thus, the actual size of grains is a time-dependent function $a+b$, where $b=N_\mathrm{ML}b_\mathrm{ML}$. Here, $N_\mathrm{ML}$ is the number of monolayers in the ice mantle.

\subsectionb{2.3}{Gas-phase chemistry}

The gas and surface reaction network was taken straightly from ALCHEMIC as the OSU\_2008\_03 gas-grain database\footnote{Available at http://www.physics.ohio-state.edu/eric/research.html}. It consists of binary reactions, direct ionization of species by cosmic rays, ionization by cosmic-ray induced photons (grain albedo 0.5), and ionization by interstellar UV photons. The rate coefficients of gas-phase reactions ($f0=f=g$) were calculated exactly as explained in Sect.2.2., \citet{Semenov10}.

\subsectionb{2.4}{Gas-grain interactions}
\label{gasgr}

Gas-grain interactions consist of charge-driven processes (from ALCHEMIC and OSU database), accretion of species on grains ($f0=g$ and $f=S$), and desorption ($f0=S$ and $f=g$). Accretion and desorption have been retained from the earlier paper \citet{Kalvans13a}, except that grains with a size of $a+b$ have been considered. Grain growth usually does not influence ice chemistry \citep{Acharyya11,Taquet12a}. The four desorption mechanisms include evaporation, whole-grain heating by heavy CRs, photodesorption by CR-induced and interstellar photons \citep{Kalvans13a}. Electron sticking to the neutral grains and the neutralization of atomic ions upon collision with a grain were retained, as described by \citet{Semenov10}, Sect.~2.3.

\subsectionb{2.5}{Physical transformations of ice}
\label{transf}

The description of the structure of the ice slightly differs from that of \citet{Kalvans13a} and \citet{Kalvans13c}, and is explained in detail below.

The ice consists of the outer surface, mantle, and cavity surface species. It was assumed that in physical and dissociation processes of ice the proportions of molecules settling into surface and cavity phases are $X_{\rm cav}=0.05$ and $X_{\rm surf}=0.2$, respectively. These numbers were adjusted to obtain a compact ice mantle, which contains around 5\% cavity species and 1\% outer surface species (at integration times longer than 0.5Myr). For surface chemical reactions (Sect.~2.6) it was assumed that there are 2,500 cavities per grain mantle with 2,000 adsorption sites each. A value of 5\% for cavity proportion in ice is approximately consistent with the experiments, showing that 5-10\% of CO can be trapped in ices during their warm-up process in protostars \citep[e.g.][]{Oberg11}.

When species stick to grains, initially a porous layer forms (outer surface phase). It is assumed that the ice is then compacted into subsurface mantle by all kinds of energetic processes. It was assumed that the full transition of species from the surface to the mantle ($f0=S, f=M$) occurs within the time of two ($N_{\rm FeCR}=2$) strikes of heavy CR particles. For cavity species, this rate is multiplied by $X_{\rm cav}$. I refer the reader to \citet{Kalvans13a} and \citet{Kalvans13c} for a more detailed discussion. 

The principal mechanism of molecule interchange between the phases is local heating by Fe-CR via the hot-soup phase. When a heavy cosmic-ray particle passes through the matter, it ionizes molecules and atoms on its path. This creates a positively-charged cylinder with a radius $R^+=(4-8.8)\times10^{-7}$cm. The emitted electrons are absorbed in the outer cylinder ($R^- \approx 3 R^+$). This system then completely neutralizes within a time-scale of $2\times10^{-12}$s \citep{Iza06}.

\citet{Leger85} have estimated that a cylinder with a radius roughly equal to $R^+$ is formed within $10^{-11}$s after the Fe-CR hit. The hot cylinder then cools and expands, until the whole grain is warmed-up. This happens within some $10^{-9}$s. However, the newer data from \citet{Iza06} point to a larger hot cylinder radius, $R_{\rm local}\approx R^-$, where the cosmic-ray energy is initially deposited. Taking these considerations into account, a cylinder with a radius of $R_{\rm local}=2\times10^{-6}$cm and a time of existence $t_{\rm local}=10^{-10}$s was taken as the standard hot-soup parameters. This time-scale is at least one order of magnitude shorter than that for evaporating (exploding) ices \citep{Cecchi10}.

In the model, it is assumed that intact molecules in the different ice phases all are transferred into a temporary `hot soup' phase with a constant rate, according to a coefficient for local heating:
   \begin{equation}
   \label{hot1}
k_{\mathrm{local}} = \sigma_{\mathrm{hot}}F_{\rm FeCR},
   \end{equation}
where $\sigma_{\rm local}=\pi R_{\rm local}^2$ is the cross section for molecule transition into the hot soup. Species return to solid ice phases with a rate coefficient
   \begin{equation}
   \label{hot2}
k_{stop}=k_{\mathrm{hot}}X_{\mathrm{freeze},f}t_{\mathrm{hot}}R_{\rm FeCR}.
   \end{equation}
$R_\mathrm{FeCR}t_{\rm local}$ is the fraction of the time spent by molecules in hot cylinder. $X_{\mathrm{freeze},f}$ is the proportion of species going to a particular phase $f$. For molecules forming the outer surface, $X_{\mathrm{freeze},S}=X_{\rm surf}$. The remainder of the molecules go either to the mantle or cavities, with $X_{\mathrm{freeze},C}/X_{\mathrm{freeze},M}=X_{\rm cav}$. No chemical reactions or desorption are assumed to occur in the locally heated cylinder.

The diffusion of H and H$_2$ is an essential part of the model. The methodology for calculating diffusion rate coefficients is given by \citet{Kalvans13a}. A temperature-dependent rate coefficient was employed in the present study, as specified in \citet{Kalvans13c}.

\subsectionb{2.6}{Ice chemistry}
\label{coldch}

The chemical model has been extensively described and analyzed in \citet{Kalvans10,Kalvans13a,Kalvans13c}. In this work, the binary reaction rate was calculated with the ``ALCHEMIC'' code \citep{Semenov10}. Reactant mobility on surfaces was calculated according to \citet{Hasegawa92}, with a binding to adsorption energy ratio $E_b/E_D=0.5$ \citep{Garrod06}. The desorption energy for molecular hydrogen was taken to be 314 K \citep{Katz99}. This model does not consider the stochastic effects associated with grain chemistry \citep{Caselli98} or reaction-diffusion competition \citep{Garrod11}. These deficiencies have not been tested for a cavity-surface diffusion process. Taking this into account, a match of calculation results and observations is not among the aims of this simulation. Rather, the general effects of molecule diffusion in a warm, porous ice are investigated.

Ice molecule dissociation by interstellar and cosmic-ray induced photons was included with the rate constants given in OSU\_2003\_03 reaction database. For subsurface species (cavity and mantle phases), the dissociation yield was adjusted for photon attenuation by overlying layers. It was assumed that, for species in the mantle, photodissociation efficiency is 0.05; i.e., only 5\% of the dissociation products do not recombine to form the parent molecule. Such a low value does not truly reflect the photoprocessing rate of ice. Rather, this is an approximation taking into account that the model does not consider reactions directly between the species in ice lattice \citep[unlike, e.g.][]{Garrod13a}. For cavity and outer surface species, photodissociation efficiency was assumed 1.0. The details on photodissociation for cavity-based mantle chemistry can be found in \citet{Kalvans13a}.

\subsectionb{2.7}{Surface-cavity diffusion}
\label{cavdif}

The main research subject in this paper was thermal migration of species between the surface and cavity phases in warm ice. This was based on the presumption that ice at 70K becomes unsteady, and the desorption and movement of rather large molecules is possible, which leaves existing pores accessible for mobile species. While this is not true for water with $E_b=2850$K (70K and  10$^{-5}$s), CO, CO$_2$, and other molecules with lower binding energies certainly are sufficiently mobile. The rate of this process is
   \begin{equation}
   \label{diss3}
k_{\rm warm-diff.} = \nu_0 \mathrm{exp}(-E_b/T(\mathrm{warm})) R_{\mathrm{FeCR}} t_{\mathrm{cool},T(\mathrm{warm})} N_{\rm hop},
   \end{equation}
where $\nu_0$ is the characteristic vibrational frequency for the molecule in consideration. $N_{hop}=200$ is the assumed number of steps required for molecules to eventually emerge to surface from a cavity or to find a cavity opening and delve into it (for surface species). This process is a deviation from previous models by \citeauthor{Kalvans10}, because it takes into account the connection of cavities to the surface, while still regarding them as separate phases. Cavity openings can be regarded as good adsorption sites and they can be easily blocked at 10K by common species, such as H$_2$O, NH$_3$, or CO$_2$. This means that in ices at thermal equilibrium ($T=10$K) the cavities are assumed to be closed. They open when the grain experiences a heating event. The exchange of molecules between the surface and the pores in ice at elevated temperatures has been supported by various experiments \citep{Sandford90, Collings03, Collings04, Fayolle11}.

Thermal migration on warm ice surfaces for H and H$_2$ molecules in the direction from surface to cavities has not been considered. This is because these species evaporate too easily at 70K.

\sectionb{3}{RESULTS}
\label{res}

\subsectionb{3.1}{Inert mantle model}
\label{r-rigid}
\begin{figure}
\centering
  \vspace{-4cm}
  \includegraphics[width=8.0cm]{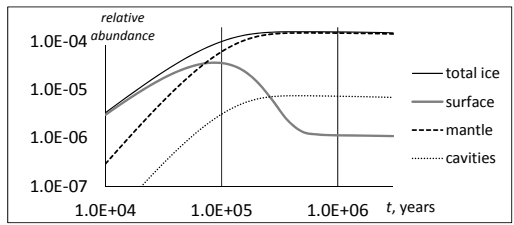}
  \vspace{-4cm}
  \caption{The relative abundance of water ice, distributed in surface, compact ice mantle, and cavity phases.\label{att-rig1}}
\end{figure}
The following sections describe the obtained results -- relative abundances of icy species. In order to avoid errors caused by the assumed and poorly-known parameters in the model, the results for the CR-induced surface-cavity diffusion are compared to a model without this process. 1Myr is used as a standard model integration time. For illustrative purposes, the calculations have been run with \textit{t}=3Myr.

The simplest case is a model without binary reactions in cavities and without hydrogen diffusion in ice. The full physical ice description, outer surface reactions, and UV and CR-photon induced dissociation of ice (\textit{S}, \textit{C}, and \textit{M}) are included, only. Fig.~\ref{att-rig1} shows the abundance of water in the three phases, in order to provide an impression on ice structure as it is reflected by the model.

At 1Myr, the basic constituents of ice are H$_2$O and CO, which are determined by the surface chemistry. Ice contains radicals, mainly H (3\% relative to water) and OH (2\%).  The radicals are produced by photodissociation and they accumulate in the mantle. The abundance of OH and H is more than a magnitude lower if photodissociation is neglected. These results are not very relevant for the ice chemistry, because of the unreasonably high radical content. The abundance of radicals is reduced by hydrogen diffusion and subsurface mantle processes (see below). To illustrate the relative proportions of the phases, Figure~\ref{att-rig1} shows the relative abundances for water in the surface, mantle, and cavities. A similar picture is relevant for all other model cases.

\subsectionb{3.2}{Reference model}
\label{r-cold}

The mantle chemistry model includes binary reactions in cavities (Sect.~2.6) and hydrogen diffusion through the ice (Sect.~2.5). It is regarded as the reference case for this paper. It is a follow-up of the models established by \citet{Kalvans10,Kalvans13a}.
\begin{figure}
  \vspace{-1cm}
  \hspace{-1.5cm}
  \includegraphics[width=12.0cm]{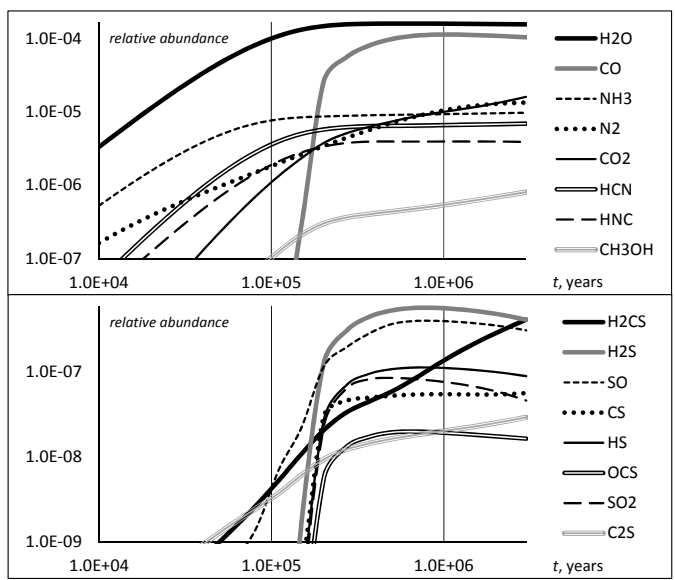}
  \vspace{-8.5cm}
  \caption{Results of the cold chemistry model. Top panel -- relative abundance of major ice species. Bottom panel -- relative abundance of sulfur species in ice.\label{att-cold1}}
\end{figure}

\subsubsectionb{3.2.1}{Major species}

Top panel of Fig.~\ref{att-cold1} shows the abundances of major ice species. An important result is that the hydrogen diffusion process diminishes the relative abundance of atomic H (compared to the inert mantle model) to less than $10^{-10}$, thus effectively reducing radical accumulation in the mantle. The abundance of heavy radicals (mainly OH, NH$_2$, CN) is two times lower and their total abundance is around 1\%. They have been turned mostly into water, ammonia, and hydrogen cyanide. Major ice species include H$_2$O, CO, N$_2$, CO$_2$, NH$_3$, and HCN with an abundance ratio of 100:71:7:6:6:4, respectively. For HNC, NH$_2$CN, OH, and HNCO the abundance ratio with water is 100:2:2:2:1, respectively.

The parameters chosen for this study ($X_{\rm cav}$, $Y_{\mathrm{dis},M}$, $N_{\rm FeCR}$) ensure a relatively rapid chemical and physical processing of the ice. They were chosen so, partially because only models with a chemically processed ice are consistent with observations of prestellar and protostellar cores. The second reason is that processing of ice by cosmic-rays, included in a physically consistent way, means a rather intense ice processing (phase changes for molecules), too.

Radicals are produced by CR-induced photons and accumulate in the subsurface ice. Reactions in the cavities are unable to completely consume these radicals.

\subsubsectionb{3.2.2}{Minor species}

The chemistry of nitrogen is dominated by molecular nitrogen, ammonia, and cyanides. HCN:HNC abundance ratio is 1.7:1 at 1Myr and is growing with time. HNCO, which is the second nitrogen molecule detected in interstellar ices (as OCN$^-$ or ``XCN'') \citep{Oberg11} along with ammonia, is reproduced relatively well.

Sulfur is distributed among H$_2$S, SO, H$_2$CS, SO$_2$, S, and CS (Fig~\ref{att-cold1}). OCS and SO$_2$ are the sulfur species most intensively produced in the mantle. The mantle-to-surface abundance ratio of around 2000 for these species is the highest among all stable molecules. This is in accordance with the results by \citet{Kalvans10}.

Cavity chemistry does not favor complex organic molecules, such species produced mostly on the surface. Among species with three or more heavy atoms, the highest abundances in ice are reached by formamide (fractional abundance $<10^{-7}$) and formic acid ($<10^{-8}$).

\subsubsectionb{3.2.3}{Comparison of mantle chemistry methods}

The model of the present study employs a different approach on subsurface mantle chemistry than the other published rate-equations model with chemical reactions in bulk ice. \citet{Garrod13a} describes subsurface chemistry with a method, similar to that for surface reactions. The main difference is that molecules in the mantle have higher binding energies and move slower. This causes reaction rates to be lower. Additionally, \citet{Garrod13a} modify binary reaction rate coefficients with reaction-diffusion competition \citep{Garrod11}. This approach significantly facilitates reactions with energy barriers. \citeauthor{Garrod13a} considers a collapsing prestellar cloud that undergoes subsequent heating by a protostar, while the present study focuses on quiescent clouds. Finally, there are several important differences in the reaction networks employed in both studies, e.g., the barrier for the CO+O reaction, the adsorption energy of atomic H, and the $E_b/E_D$ ratio.

The abovementioned differences imply that a direct comparison of the two methods for mantle chemistry description would not be productive. The physical conditions and calculated rate coefficients for many reactions are different. However, both models have the common result that subsurface chemistry facilitates the formation of complex species, including COMs. Such synthesis is often promoted by the inability for radicals in ice to be rapidly hydrogenated in the H-poor subsurface conditions. Reactions in the cavities seem to be essential also for sulfur chemistry \citep[Sect.~3.2.2 and][]{Kalvans10}, however \citet{Garrod13a} does not analyze the chemistry of sulfur.

It is important to note that the cavity reaction approach of and bulk-ice reaction approach are complementary, not contradicting methods. It has been demonstrated by a rigorous off-lattice Monte-Carlo model \citet{Garrod13b} that porosity is an intrinsic feature of interstellar ices. This may well mean that processes related to both, cavities and bulk ice, have to be considered for rate-equation-based models that intend to fully describe interstellar ices.

\subsectionb{3.3}{Surface-cavity migration of species}
\label{r-warm-migr}
\begin{figure}
\centering
  \vspace{-9.5cm}
  \includegraphics[width=14.0cm]{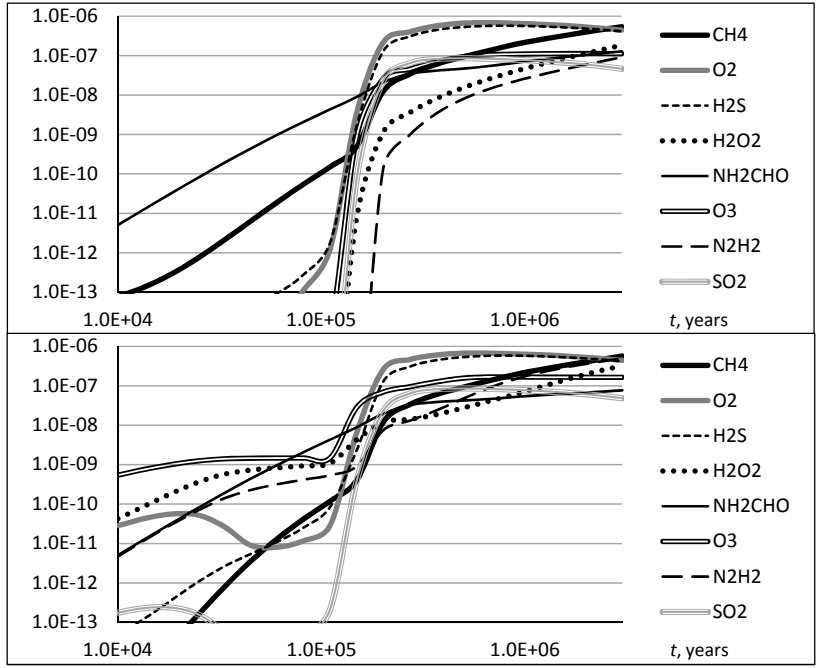}
  \vspace{-2cm}
  \caption{Comparison of relative abundances for selected species. Top panel --reference model, bottom panel -- model with CR-induced surface-cavity diffusion.\label{att-sal1}}
	\end{figure}
\begin{figure}
\centering
  \vspace{-2.3cm}
  \includegraphics[width=14.0cm]{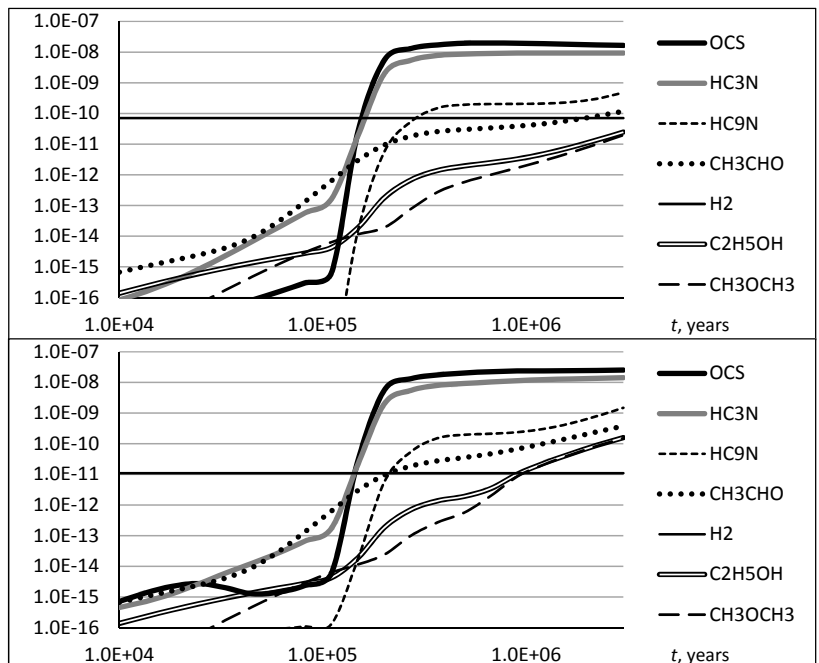}
  \vspace{-10cm}
  \caption{Comparison of relative abundances for selected species. Top panel --reference model, bottom panel -- model with CR-induced surface-cavity diffusion. Note that the ice abundance of H$_2$ is a constant.\label{att-sal2}}
\end{figure}
Figures \ref{att-sal1} and \ref{att-sal2} show relative abundances for important species that experience significant change in their abundance, relative to the reference model.

Migration-induced phase transition between the surface and the cavities of warm species causes a reduction of H$_2$ content in the mantle, and a supply of mantle-generated radicals available for reactions with stable surface molecules. This gives rise to reaction chains that produce radicals of different types, which subsequently affect a wide array of species. Table~1 shows that at 1Myr the species whose abundances have been greatly enhanced by the warm diffusion process are hydrazine N$_2$H$_2$, ethyl alcohol C$_2$H$_5$OH, dimethyl ether CH$_3$OCH$_3$, acetaldehyde CH$_3$CHO, hydrogenated C-N compounds, carbonyl sulfide OCS, hydrogen peroxide H$_2$O$_2$, ozone O$_3$, and cyanopolyynes. These changes happen at the expense of a small abundance decrease for important ice species CO$_2$, NH$_3$, N$_2$, HCN, O$_2$, H$_2$CS.

The alterations are associated with two processes: (1) with the possibility for radicals, which have been accumulated in ice, to reach the outer surface, where they quickly react, and, (2) with efficient hydrogenation of cavity species thanks to a significantly faster hydrogen diffusion rate through ice at elevated temperature.
\begin{table}[!t]
\begin{center}
\vbox{\footnotesize\tabcolsep=3pt
\parbox[c]{124mm}{\baselineskip=10pt
{\smallbf\ \ Table 1.}{\small\
The abundance of chemical species, relative to hydrogen, calculated with the CR-induced diffusion model at 1Myr. Also given is the abundance change ($\pm$) in percentage, respective to the reference model.\lstrut}}
\label{tab-res}
\begin{tabular}{lcr|lcr|lcr|lcr}
\hline
Species & Abund. & $\pm$,\% & Species & Abund. & $\pm$,\% & Species & Abund. & $\pm$,\% & Species & Abund. & $\pm$,\% \\
\hline
C & 5.9E-08 & -1 & C$_7$ & 1.7E-12 & 33 & H$_2$C$_3$N & 1.1E-09 & -1 & N$_2$O & 4.9E-08 & 1 \\
C$_2$ & 4.2E-08 & -2 & C$_7$H & 4.9E-11 & 7 & H$_2$C$_3$O & 7.0E-10 & 4 & NH & 2.1E-07 & 3 \\
C$_2$H & 1.8E-09 & -5 & C$_7$H$_2$ & 2.5E-09 & 4 & H$_2$C$_5$N & 2.5E-10 & 12 & NH$_2$ & 6.0E-07 & 2 \\
C$_2$H$_2$ & 4.1E-10 & -4 & C$_7$H$_3$ & 7.6E-11 & 9 & H$_2$C$_7$N & 3.6E-11 & 14 & NH$_2$CHO & 7.8E-08 & 11 \\
C$_2$H$_2$N & 4.6E-09 & 240 & C$_7$H$_4$ & 1.2E-09 & -1 & H$_2$C$_9$N & 1.0E-11 & 18 & NH$_2$CN & 3.0E-06 & 0 \\
C$_2$H$_2$O & 3.5E-08 & 2 & C$_7$N & 6.0E-11 & 10 & H$_2$CN & 5.4E-11 & 11 & NH$_3$ & 9.3E-06 & -1 \\
C$_2$H$_3$ & 3.9E-10 & 7 & C$_8$ & 2.6E-12 & 37 & H$_2$CO & 6.9E-07 & 8 & NO & 1.3E-07 & -2 \\
C$_2$H$_3$N & 4.6E-08 & 16 & C$_8$H & 4.2E-11 & 8 & H$_2$CS & 1.2E-07 & -13 & NO$_2$ & 1.2E-08 & 0 \\
C$_2$H$_4$ & 5.2E-09 & 0 & C$_8$H$_2$ & 2.0E-09 & 4 & H$_2$O & 1.6E-04 & 0 & NS & 5.6E-08 & -4 \\
C$_2$H$_4$O & 5.6E-11 & 39 & C$_8$H$_3$ & 5.8E-11 & 9 & H$_2$O$_2$ & 6.7E-08 & 43 & O & 2.1E-07 & 1 \\
C$_2$H$_5$ & 5.8E-10 & 9 & C$_8$H$_4$ & 1.1E-09 & -4 & H$_2$S & 5.7E-07 & 2 & O$_2$ & 6.1E-07 & -5 \\
C$_2$H$_5$OH & 1.8E-11 & 400 & C$_9$ & 1.7E-10 & 7 & H$_2$S$_2$ & 3.4E-10 & -18 & O$_2$H & 1.5E-08 & 209 \\
C$_2$H$_6$ & 1.5E-09 & 2 & C$_9$H & 3.0E-11 & 30 & H$_2$SIO & 3.1E-08 & 0 & O$_3$ & 9.7E-08 & -12 \\
C$_2$N & 1.9E-09 & 73 & C$_9$H$_2$ & 8.0E-10 & 5 & H$_3$C$_5$N & 5.7E-09 & 5 & OCN & 1.1E-07 & -1 \\
C$_2$S & 2.0E-08 & -2 & C$_9$H$_3$ & 2.1E-11 & 13 & H$_3$C$_7$N & 7.5E-10 & 5 & OCS & 6.5E-08 & 237 \\
C$_3$ & 8.3E-10 & 12 & C$_9$H$_4$ & 4.0E-10 & 0 & H$_3$C$_9$N & 2.2E-10 & 5 & OH & 2.2E-06 & 5 \\
C$_3$H & 1.0E-08 & -1 & C$_9$N & 1.9E-11 & 18 & H$_4$C$_3$N & 7.6E-10 & 2 & P & 7.3E-08 & 0 \\
C$_3$H$_2$ & 8.9E-08 & -2 & C$_{10}$ & 2.7E-08 & 0 & H$_5$C$_3$N & 1.6E-08 & 3 & PH & 3.5E-10 & 1 \\
C$_3$H$_3$ & 7.8E-09 & 6 & CCO & 1.0E-09 & 16 & HC$_2$NC & 3.2E-09 & 3 & PH$_2$ & 1.3E-12 & 10 \\
C$_3$H$_3$N & 7.1E-09 & 2 & CCP & 1.0E-11 & 1 & HC$_2$O & 9.4E-10 & 17 & PN & 2.1E-09 & 0 \\
C$_3$H$_4$ & 4.6E-08 & 5 & CH & 3.8E-08 & -3 & HC$_3$N & 9.7E-09 & 4 & PO & 7.0E-10 & 0 \\
C$_3$N & 2.5E-10 & 16 & CH$_2$ & 6.4E-09 & 20 & HC$_3$O & 3.6E-11 & 10 & S & 7.0E-08 & -8 \\
C$_3$O & 3.6E-11 & 11 & CH$_2$NH$_2$ & 5.2E-09 & 1161 & HC$_5$N & 4.0E-09 & 10 & S$_2$ & 7.8E-09 & -14 \\
C$_3$P & 3.9E-12 & 3 & CH$_2$O$_2$ & 7.1E-09 & 0 & HC$_7$N & 6.9E-10 & 9 & SI & 6.4E-09 & 4 \\
C$_3$S & 6.0E-09 & -4 & CH$_2$OH & 2.6E-08 & 21 & HC$_9$N & 2.3E-10 & 11 & SIC & 3.6E-09 & 0 \\
C$_4$ & 2.3E-10 & 16 & CH$_3$ & 1.6E-08 & 1 & HCCN & 4.2E-09 & 494 & SIC$_2$ & 6.5E-09 & 1 \\
C$_4$H & 8.9E-10 & 14 & CH$_3$C$_3$N & 6.1E-11 & 1 & HCCP & 4.7E-12 & 4 & SIC$_2$H & 2.2E-08 & 1 \\
C$_4$H$_2$ & 3.2E-08 & 7 & CH$_3$C$_4$H & 2.9E-10 & -1 & HCN & 6.5E-06 & -3 & SIC$_2$H$_2$ & 1.4E-10 & 0 \\
C$_4$H$_3$ & 8.3E-10 & 7 & CH$_3$C$_5$N & 5.3E-12 & 3 & HCNC$_2$ & 3.1E-10 & 8 & SIC$_3$ & 2.9E-10 & -1 \\
C$_4$H$_4$ & 1.4E-08 & -3 & CH$_3$C$_6$H & 5.2E-11 & -1 & HCO & 8.0E-08 & 6 & SIC$_3$H & 9.6E-11 & 1 \\
C$_4$N & 1.6E-08 & 4 & CH$_3$C$_7$N & 6.3E-13 & 3 & HCOOCH$_3$ & 6.5E-13 & -1 & SIC$_4$ & 5.9E-13 & 1 \\
C$_4$P & 9.3E-13 & 5 & CH$_3$N & 1.2E-08 & 701 & HCP & 6.4E-11 & 1 & SICH$_2$ & 1.7E-08 & 0 \\
C$_4$S & 3.2E-10 & -4 & CH$_3$NH & 4.1E-09 & 1053 & HCS & 9.9E-10 & -19 & SICH$_3$ & 2.4E-11 & 0 \\
C$_5$ & 1.3E-11 & 35 & CH$_3$OCH$_3$ & 1.7E-11 & 725 & HCSI & 1.5E-09 & 0 & SIH & 2.5E-09 & 12 \\
C$_5$H & 3.5E-10 & 7 & CH$_4$ & 2.5E-07 & 14 & HNC & 4.0E-06 & 1 & SIH$_2$ & 1.6E-08 & 7 \\
C$_5$H$_2$ & 1.7E-08 & 1 & CH$_4$O & 5.5E-07 & 2 & HNC$_3$ & 3.5E-09 & 9 & SIH$_3$ & 3.4E-09 & 32 \\
C$_5$H$_3$ & 6.0E-10 & 7 & CH$_5$N & 1.5E-07 & 20 & HNCO & 1.4E-06 & 2 & SIH$_4$ & 6.2E-07 & 0 \\
C$_5$H$_4$ & 8.4E-09 & 0 & CHNH & 1.2E-09 & 1667 & HNO & 6.5E-07 & -5 & SIN & 1.3E-09 & 0 \\
C$_5$N & 3.2E-10 & 10 & CN & 7.3E-07 & -2 & HNSI & 1.1E-08 & 0 & SINC & 2.6E-12 & 0 \\
C$_6$ & 5.4E-12 & 32 & CO & 1.1E-04 & -1 & HPO & 7.5E-12 & 1 & SIO & 6.4E-07 & 0 \\
C$_6$H & 1.2E-10 & 9 & CO$_2$ & 1.0E-05 & 0 & HS & 1.1E-07 & -1 & SIO$_2$ & 3.0E-08 & 1 \\
C$_6$H$_2$ & 5.7E-09 & 4 & CP & 2.5E-11 & 3 & HS$_2$ & 3.7E-10 & -18 & SIS & 4.2E-11 & -3 \\
C$_6$H$_3$ & 1.8E-10 & 9 & CS & 5.2E-08 & -6 & N & 5.4E-08 & -1 & SO & 3.9E-07 & -3 \\
C$_6$H$_4$ & 3.2E-09 & -4 & H & 5.8E-11 & -1 & N$_2$ & 1.0E-05 & -2 & SO$_2$ & 7.3E-08 & -5 \\
C$_6$H$_6$ & 7.1E-10 & 4 & H$_2$ & 7.1E-11 & 0 & N$_2$H$_2$ & 2.6E-08 & -1 &  &  &  \lstrut \\
\hline
\end{tabular}
}
\end{center}
\vskip-8mm
\end{table}
The most common radicals are OH, CN, S, NH$_2$, NH, and O. They are generated from the most abundant ice species. The abundances of many radicals are slightly lower than in the reference model. Mobile radicals (atomic species, CH$_x$) more effectively ascend to the surface, where they become exposed to atomic hydrogen accreting from the gas phase. The less mobile radicals, OH, NH$_2$, NH, and CN are hydrogenated via the second process (in above paragraph).

An important family of compounds are species whose abundance is affected by the enhanced mobility of the atomic products of CO photodissociation. These mainly include carbon chains, in the form of the stable cyanopolyynes, and CO$_2$. Another important change, relative to the reference model, is the abundance of molecular hydrogen, which is six times lower in the warm diffusion model. This is caused by hydrogen diffusion in ice (Sect.~2.5), which proceeds faster at higher temperatures.

An interesting result is the abundance increase for several complex organic molecules (COM). Most notably, these include dimethyl ether, ethyl alcohol, and acetaldehyde. This is because the formation reactions of these species involve the methyl radical CH$_3$, which is generated in significant amounts in the mantle by the dissociation of methane and methanol. The low binding energy of CH$_3$ means that it becomes extremely mobile at 70K, and can effectively ascend to the surface and react. The abundance of CH$_4$ and CH$_3$OH is increased too, but the effect on COMs exceeds two orders of magnitude, because they have inherently low abundances in the reference model. The abundances of most other COMs have remained unchanged.

\sectionb{4}{CONCLUSION}

Pores in ice may act as a transport route for molecules between outer surface and the subsurface mantle. This process can become efficient in ices that have been warmed-up by a heavy cosmic-ray hit. Molecule migration in 70K ice, warmed for 10$^{-5}$s, supplies light radicals to the outer surface. The radicals accumulate in the mantle because of cosmic-ray induced photodissociation. Additionally, enhanced diffusion of hydrogen in warmed-up ice increases the rate of hydrogenation of molecules and radicals. 

These processes result in an abundance increase for several species thought to be associated with interstellar ices, and observed in prestellar cores and young stellar objects. Most notably, these include COMs \citep{Bacmann12}, carbonyl sulfide \citep{Palumbo97}, hydrogen peroxide and hydroperoxyl radical O$_2$H \citep{Bergman11,Parise12}. This coincidence may point out either to CR-induced processing of interstellar ices, or efficient coupling between the surface and the mantle, as indicated also by \citet{Garrod13a}. The gas phase abundances of these species are not significantly different in the warm diffusion model and the reference model.

\thanks{I thank Dmitry Semenov for providing the ALCHEMIC astrochemistry code. This research has made use of NASA's Astrophysics Data System.}

\bibliography{wdif}
\bibliographystyle{ba}

\end{document}